# High-fidelity quantum logic operations using linear optical elements

J. D. Franson, M. M. Donegan, M. J. Fitch, B. C. Jacobs, and T. B. Pittman

Johns Hopkins University

Applied Physics Laboratory

Laurel, MD 20723

Abstract:

Knill, Laflamme, and Milburn [Nature **409**, 46 (2001)] have shown that quantum logic operations can be performed using linear optical elements and additional ancilla photons. Their approach is probabilistic in the sense that the logic devices fail to produce an output with a failure rate that scales as $1/n$, where $n$ is the number of ancilla. Here we present an alternative approach in which the logic devices always produce an output with an intrinsic error rate that scales as $1/n^2$, which may have several advantages in quantum computing applications.

In an earlier paper [1], Knill, Laflamme, and Milburn (KLM) proposed a method for implementing probabilistic quantum logic gates using linear optical elements, additional ancilla photons, and post-selection based on the results of measurements made on the ancilla. The output of these devices is known to be correct (in principle) when certain measurements results are obtained, but no output is produced for the remaining events, which occur with a failure rate that scales as $1/n$ where $n$ is the number of ancilla. Here we propose an alternative approach in which the logic devices always produce an output with an intrinsic error rate that scales as $1/n^2$, which is expected to have several advantages in quantum computing applications.

Logic operations are inherently nonlinear, whereas nonlinear interactions at the single-photon level are difficult to achieve because of the low electric field associated with a single photon. The KLM approach makes use of the fact that the measurement process itself is nonlinear. For example, a single-photon detector either registers a photon or not, which is a nonlinear response to the field. By accepting only those events in which measurements made on the ancilla yield a specified result, the system can be left in a post-selected final state that corresponds to the desired quantum logic operation. Depending on the results of the measurements, it may be necessary to apply a pre-determined classical correction to the output states in order to obtain the desired logical output. We have experimentally demonstrated several quantum logic devices of this kind [2, 3] that have a probability of success of $1/2$.

Quantum error correction requires relatively small error rates, however, and KLM showed that the failure probability could decrease as $1/n$ in the limit of large $n$. Our approach is motivated by the fact that quantum error correction automatically identifies the presence of an error via the appropriate syndrome [4], in which case there may be little practical advantage in independently knowing that a logic operation has succeeded based on measurements made on the ancilla. In addition, even when the KLM approach indicates that a logic operation has succeeded, there may still be errors from technical sources, such as photon absorption or decoherence. As a result, our strategy is to accept the output of all of the logic operations while choosing the initial state of the ancilla photons to minimize the overall error rate, which is equivalent to maximizing the fidelity of the output. The ancilla will still be prepared using post-selection, but the calculations themselves will be performed with the maximum fidelity. We will refer to our approach as the high-fidelity or "Hi-Fi" approach in order to distinguish it from the original KLM method; the two approaches differ only in their overall strategy and the choice of the entangled ancilla state.

Gottesman and Chuang [5] showed that quantum logic operations could be performed using quantum teleportation, where the desired logic operation is applied to the entangled pair of ancilla rather than to the input qubits. Probabilistic techniques [1-3, 6-8] can be applied repeatedly until the necessary ancilla state has been generated, after which the logic operation can be successfully performed by teleporting the qubits of interest. As a result, we begin by describing a quantum teleportation protocol with an error rate that scales as $2/n^2$.

The input qubit $q$ will be represented by a single optical mode, such as a single-mode optical fiber. The absence of a photon will represent a logical value of 0, while the presence of a single photon will represent a logical value of 1. The input qubit to be teleported is assumed to be in an arbitrary initial state $|q\rangle = \alpha_0 |0\rangle + \alpha_1 |1\rangle$. Unlike the original quantum teleportation protocol [9], our approach makes use of $n$ ancilla photons distributed among two sets of modes, which we will refer to as the x and y registers, as illustrated in the upper part of Figure 1. Each of these registers contains n modes and their states will be represented by $|x\rangle$ and $|y\rangle$. The modes in register x will be labeled by index $l$ with values from 1 to $n$ while the modes in register y will be labeled by index $l$ with values from $n+1$ to $2n$. The mode containing the initial qubit will be labeled $l=0$.

Our use of the x and y ancilla registers is the same as in the KLM approach [1], except that we assume a more general form for the initial entangled ancilla state $|\psi_{A0}\rangle$ given by

$$|\psi_{A0}\rangle = \sum_{j=0}^{n} f(j) \; |1\rangle^{j} |0\rangle^{n-j} \; |0\rangle^{j} |1\rangle^{n-j} \qquad (1)$$

Here the $f(j)$ are arbitrary coefficients whose values will be chosen to maximize the fidelity. The contents of modes 1 through $2n$ are listed from left to right, where $|1\rangle^{j}$ indicates that there is one photon in the first $j$ modes, $|0\rangle^{n-j}$ indicates that there are no photons in the next $n-j$ modes, etc. We have left a space in Eq. (1) to separate the states of the x and y registers. It can be seen that $|\psi_{A0}\rangle$ contains exactly n ancilla photons and that the y register is the same as the x register except that all of its qubits have been flipped, as illustrated in Figure 1 for the case of $n=5$ and $j=2$.

The first step in the teleportation process is to apply a Fourier transform $\hat{F}$ to the combination of $q$ and the $x$ register, which we will refer to as register $c$; i.e., $|c\rangle = |q\rangle |x\rangle$. The Fourier transform corresponds to the operator transformation

$$\hat{a}_l^\dagger \rightarrow \frac{1}{\sqrt{n+1}} \sum_{p=0}^{n} e^{2\pi i p l/(n+1)} \hat{a}_p^\dagger \qquad (2)$$

where $i = \sqrt{-1}$. $\hat{F}$ can be implemented efficiently using a set of beam splitters and phase shifters [10, 11]. After the Fourier transform, the numbers $r_l$ of photons in modes 0 through n are measured and the total number of photons will be denoted by $k = \sum r_l$.

Consider the case in which the measurements yield a particular value of k. The measurement process is then a projection onto the subspace $|S_k\rangle$ of the original state vector that contains a total of k photons:

$$|S_k\rangle = \alpha_0 |c_0\rangle |y_0\rangle + \alpha_1 |c_1\rangle |y_1\rangle \qquad (3)$$

where

$$|c_0\rangle = f(k)\hat{F}|0\rangle |1\rangle^k |0\rangle^{n-k}$$

$$|y_0\rangle = |0\rangle^k |1\rangle^{n-k}$$

$$|c_1\rangle = f(k-1)\hat{F}|1\rangle^1 |1\rangle^{k-1} |0\rangle^{n-k+1} \quad (4)$$

$$|y_1\rangle = |0\rangle^{k-1} |1\rangle^{n-k+1}$$

Here $|c_0\rangle$ corresponds to the case in which no photons were initially present in $q$ and $k$ photons were present in $x$, while $|c_1\rangle$ corresponds to the case in which there was one photon initially present in $q$ and $k-1$ photons present in x. We use the convention that $f(j)=0$ unless $0 \leq j \leq n$, which eliminates one of the terms in Eqs. (3) and (4) if $k=0$ or $k=n+1$.

As noted by KLM, applying a phase shift of $\exp[2\pi i l r_l/(n+1)]$ to each mode $l$ in register $c$ after applying $\hat{F}$ is equivalent to shifting modes 0 through $n$ right by one location before applying $\hat{F}$. This is illustrated in Fig. 2. As a result, the relevant terms in $|c_0\rangle$ and $|c_1\rangle$ differ only by a phase factor given by

$$\Delta\varphi = \prod_{l=0}^{n} e^{2\pi i l r_l/(n+1)} \quad (5)$$

aside from the different factors of $f(k)$. This phase factor can be compensated by applying a classical correction using Pockels cells, for example. After the phase correction, the projective measurement leaves the system in the state $|\psi_k\rangle$ given by

$$|\psi_k\rangle = c_n(\alpha_0 f(k)|y_0\rangle + \alpha_1 f(k-1)|y_1\rangle)|M\rangle \quad (6)$$

Here $c_n$ is a normalization constant and $|M\rangle$ is the final state of the measurement device, which is the same for the $\alpha_0$ and $\alpha_1$ terms.

The contents of mode $n+k$ in the $|y\rangle$ register is now selected as the output. It can be seen from the arrow in Figure 1 that this qubit will have the correct value of 0 or 1 and that the remaining modes of register y will be in the same state $|y_R\rangle$ for both terms in Eq. (6). The final output state $|\psi_{out}\rangle$ from the teleportation is thus given by

$$|\psi_{out}> = c_n(\alpha_0 f(k)|0> + \alpha_1 f(k-1)|1>)|M>|y_R> \qquad (7)$$

The states $|M>$ and $|y_R>$ are common to both terms and have no effect on any subsequent computations.

It can be seen that Eq. (7) would correspond to the correctly teleported state if $f(k)=f(k-1)$. In the KLM approach, all of these coefficients are taken to be equal. In that case the teleportation succeeds with certainty unless $k=0$ or $k=n+1$, in which case the input state has been determined and the output of the logic operation is rejected. This occurs with probability $1/n$ in the limit of large $n$ and gives a correspondingly large failure rate. In our approach, the output is accepted in all cases and we choose the values of the $f(j)$ to minimize the average error rate. The probability amplitude for the output to be in the correct state is the projection $<q|\psi_{out}>$, and the corresponding probability $P_S$ of obtaining a correct output is

$$P_S = |<q|\psi_{out}>|^2 \qquad (8)$$

The probability that an error would be detected by an error correction algorithm, for example, is $P_E = 1 - P_S$, and $P_S$ is the square of the fidelity [4] of the output.

The intuitive idea behind our approach is that it would be better to evenly distribute the probability amplitude for an error over all of the terms in the entangled ancilla state rather than concentrating it on just a few ( $k=0$ or $k=n+1$). Since the error probability depends on the squares of the amplitudes, one might expect to reduce the overall error by a factor of $1/n$ by distributing the errors evenly. The expected error reduction is roughly analogous to the quantum Zeno effect [12].

With that in mind, we assume that $f(j)$ is a slowly-varying function in the limit of large $n$, in which case the approximation

$$f(k-1) = f(k) - \frac{df}{dk} \qquad (9)$$

holds, where the derivative has been multiplied by $\delta k = -1$. We also define a parameter $\varepsilon$ by writing

$$\frac{df}{dk} = f'(k) = \varepsilon(k) f(k) \qquad (10)$$

and we note that $\varepsilon$ can be made small in the limit of large $n$, since $k$ will take on large values. Evaluating the normalization constant $c_n$ and taking the projection in Eq. (8) gives an expression

for $P_S$:

$$P_S = \left[ \frac{|\alpha_0|^2 f(k) + |\alpha_1|^2 f(k-1)}{(|\alpha_0|^2 f(k)^2 + |\alpha_1|^2 f(k-1)^2)^{1/2}} \right]^2 \qquad (11)$$

Making use of Eqs. (9) and (10) allows this to be rewritten as

$$P_S = \frac{[|\alpha_0|^2 + |\alpha_1|^2(1-\varepsilon)]^2}{|\alpha_0|^2 + |\alpha_1|^2(1-\varepsilon)^2} \qquad (12)$$

Expanding to second order in $\varepsilon$ gives

$$P_S = 1 - |\alpha_0|^2 |\alpha_1|^2 \varepsilon^2 = 1 - P_0 P_1 \left( \frac{f'(k)}{f(k)} \right)^2 \qquad (13)$$

where we have defined $P_0 = |\alpha_0|^2$ and $P_1 = |\alpha_1|^2$, and used the definition of $\varepsilon$.

If we assume that $P_0$ and $P_1$ are uniformly distributed between 0 and 1, then the average value of $P_0 P_1$ is equal to $1/6$. In the limit of large $n$, the probability $P(k)$ of obtaining $k$ photons is $f(k)^2$, in which case the average value of $\varepsilon^2$ is given by

$$\overline{\varepsilon^2} = \int_0^n P(k)\varepsilon^2 dk = \int_0^n f'(k)^2 dk \qquad (14)$$

Combining Eqs. (13) and (14) gives an average error rate of

$$P_E = \frac{1}{6} \int_0^n f'(k)^2 dk \qquad (n \gg 1) \qquad (15)$$

In order to evaluate the average error rate, we will consider the simple case in which $f(j)$ is zero for $j=0$ or $j=n$ and increases linearly to a maximum value at $j=n/2$ as determined by the normalization requirement that $\sum f(j)^2 = 1$. This choice of coefficients is illustrated in Fig. 3 for the case of $n=10$. The probability of a large error near $k=0$ or $k=n+1$ becomes negligibly small in the limit of large $n$ and the probability amplitude for an error ($f'$) is evenly distributed across all of the terms in the entangled ancilla state. Evaluating the integral in Eq. (14) for this choice of $f(k)$ gives

$$P_E = 2/n^2 \qquad (n \gg 1) \tag{16}$$

The optimal choice of $f(k)$ can be found numerically and gives an error rate 18% less than that above.

Having described a high-fidelity approach for quantum teleportation, we can now consider the implementation of two-qubit quantum logic gates as originally suggested by Gottesman and Chuang [5]. We will take $q$ to be the control qubit and introduce a second qubit $q'$ as the target, where $q'$ is initially in the arbitrary state $|q'\rangle = \alpha_0'|0'\rangle + \alpha_1'|1'\rangle$. As illustrated in Fig. 1, $q'$ will have associated ancilla registers $x'$ and $y'$ that will be used in its teleportation along with the teleportation of $q$ using registers $x$ and $y$ as described above. The goal will be to apply the desired logic operation to the ancilla in registers $y$ and $y'$ before the teleportation in order to implement the desired logic operations on $q$ and $q'$.

We first consider the implementation of a controlled sign flip [1] in which the sign of the $\alpha_1'|1'\rangle$ term is to be flipped if $q=1$. In order to implement this operation, we take the two sets of ancilla to be in an entangled state $|\psi_{AA'}\rangle$ given by

$$|\psi_{AA'}\rangle = \sum_{j=0}^{n} f(j) \; |1\rangle^j |0\rangle^{n-j} \; |0\rangle^j |1\rangle^{n-j} \sum_{j'=0}^{n} (-1)^{jj'} f(j') |1\rangle^{j'} |0\rangle^{n-j'} \; |0\rangle^{j'} |1\rangle^{n-j'} \tag{17}$$

which is similar to the entangled state used by KLM except for the factors of $f(j)$ and $f(j')$. This state would correspond to the tensor product of two sets of ancilla in the initial state of Eq. (1) except that the factor of $(-1)^{jj'}$ entangles the two. Here we perform separate Fourier transforms $\hat{F}$ and $\hat{F}'$ on the two combined registers $c$ and $c'$ and we measure the numbers $r_l$ and $r_{l'}$ of photons in each mode, with totals denoted $k$ and $k'$. After the same phase corrections described above, the measurement projects the system into the state $|\psi_C\rangle$ given by

$$|\psi_C\rangle = c_n(-1)^{kk'}[\alpha_0\alpha_0' f(k)f(k')|0\rangle|0'\rangle + (-1)^{-k}\alpha_0\alpha_1' f(k)f(k'-1)|0\rangle|1'\rangle$$
$$+ (-1)^{-k'}\alpha_1\alpha_0' f(k-1)f(k')|1\rangle|0'\rangle + (-1)^{-k-k'+1}\alpha_1\alpha_1' f(k-1)f(k'-1)|1\rangle|1'\rangle] \tag{18}$$
$$\times |y_R\rangle|y_R'\rangle|M\rangle|M'\rangle$$

This state differs from the desired output state by various sign factors, which can be corrected as follows: If both $k$ and $k'$ are even, no further correction is required. If $k$ is even and $k'$ is odd, we apply a sign flip (180° phase shift) to the $|q\rangle=1$ components, which can be accomplished using a Pockels cell, for example. If $k$ is odd and $k'$ is even, we apply a sign flip to the $|q'\rangle=1$ components. If both $k$ and $k'$ are odd, we apply a sign flip to both the $|q\rangle=1$ and $|q'\rangle=1$ components. With these corrections, the system is left in a final state given by

$$|\psi_{out}\rangle = c_n[\alpha_0\alpha_0' f(k)f(k')|0\rangle|0'\rangle + \alpha_0\alpha_1' f(k)f(k'-1)|0\rangle|1'\rangle$$
$$+\alpha_1\alpha_0' f(k-1)f(k')|1\rangle|0'\rangle - \alpha_1\alpha_1' f(k-1)f(k'-1)|1\rangle|1'\rangle]|y_R\rangle|y_R'\rangle|M\rangle|M'\rangle \quad (19)$$

which corresponds to the desired result aside from the factors of $f$. The total error rate can be shown to be the sum of the error rates from the two teleportations in the limit of large $n$, giving $P_E = 4/n^2$ for the choice of coefficients shown in Fig. 3. The optimal choice of $f(j)$ gives an error rate that is 32% less.

A controlled-NOT gate can be implemented by applying simple single-qubit transformations before and after the controlled sign flip gate. It is interesting, however, to consider the possibility of directly implementing a CNOT gate by using an entangled state of the form

$$|\psi_{AA'}\rangle = \sum_{j=0}^{n} f(j) |1\rangle^j |0\rangle^{n-j} |0'\rangle^j |1'\rangle^{n-j} \sum_{j'=0}^{n} \prod_{ll'} \hat{C}_{ll'} f(j')|1\rangle^{j'}|0\rangle^{n-j'} |0'\rangle^{j'}|1'\rangle^{n-j'} \quad (20)$$

Here the operator $\hat{C}_{ll'}$ applies a CNOT operation between modes $l$ and $l'$ in the $y$ and $y'$ registers, with $y$ the control and $y'$ the target. Although this method does give the correct output values for $q$ and $q'$, the various terms are entangled with $|y_R'\rangle$ as well as with $|y_R\rangle$, where $|y_R'\rangle$ corresponds to the remaining qubits in register $y'$ with the values of the qubits flipped. This entanglement destroys any interference effects in any subsequent calculations and is unacceptable for quantum computing applications. Thus it appears that the original method suggested by Gottesman and Chuang cannot be readily extended to arbitrary quantum logic gates using large numbers of ancillas, and it may be limited to the controlled sign flip operation [1] for this reason.

It is interesting to note that our approach does not rely on post-selection in the sense that all of the logic operations are accepted after deterministic corrections have been applied. The nonlinearity still originates in the reduction of the state vector during the measurement process, but the process might be better referred to as post-correction rather than post-selection.

We have seen that our "Hi-Fi" approach can give average error rates for quantum logic operations that scale as $4/n^2$, as compared to the $2/n$ scaling for the failure rate in the original KLM approach. As we mentioned before, there may be little advantage in having an independent indication of the failure of a logic operation in quantum computing applications, since quantum error correction algorithms can already identify and correct any errors. The most important consideration is that the total error rate must be below the threshold for fault-tolerant error correction [4]. Our approach allows the same error tolerance to be achieved using many fewer ancilla photons in the limit of large n. This has the obvious advantage of reducing the resources required to generate the entangled ancilla states, which may involve the use of relatively inefficient post-selection algorithms [2, 3, 7-8] as will be described in a subsequent paper. But more importantly, the rate of decoherence and photon absorption will increase linearly with the number of ancilla involved, so that the number of ancilla should be kept as small as possible in order to avoid additional sources of error and decoherence. As a result, we expect that our approach may have practical advantages in the implementation of a quantum computer.

This work was supported by the Office of Naval Research and by Independent Research and Development funds.

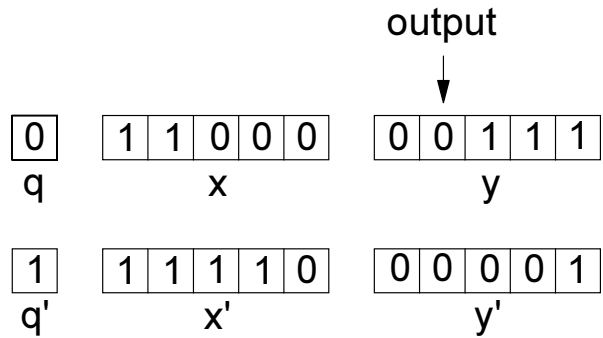

Fig. 1 Ancilla registers $x$, $y$, $x'$, and $y'$ used in the quantum teleportation of two input qubits $q$ and $q'$ [1]. The example shown here corresponds to $n=5$ and the values of the qubits correspond to one term in the entangled state where $q=0$, $j=2$, $q'=1$, and $j'=4$. If a total of $k=2$ photons are found in $q$ and $x$ after the Fourier transform, for example, then the output of the teleported state for $q$ would be taken from mode $l=n+k$, as indicated by the arrow. It can be seen that this corresponds to the correct value of $q=0$. If a total of two photons were found but the input value of $q$ was equal to 1, that would correspond to $j=1$ and the output qubit at that location would have had the value 1, as required in that case. The projective nature of the measurement process gives the appropriate superposition of these two output values.

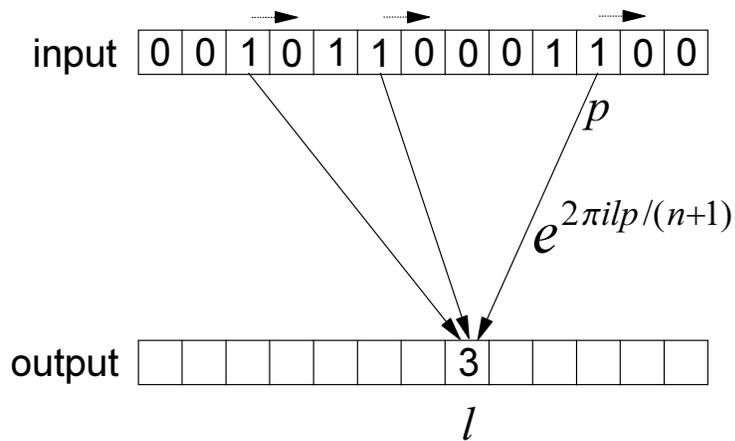

Fig. 2  Translational property of the Fourier transform $\hat{F}$. In this example, 3 photons are present in mode $l$ in the output state. There is some probability amplitude that the three photons originated in the three input modes indicated by the solid arrows, with an associated phase shift for each as shown in the figure. If the input state is translated one mode to the right, as indicated by the dashed arrows, the same probability amplitude will occur but with a phase shift $\Delta\varphi$ as given in the text. This translational property is responsible for the coherence of the $\alpha_0$ and $\alpha_1$ terms in the quantum teleportation of Fig. 1.

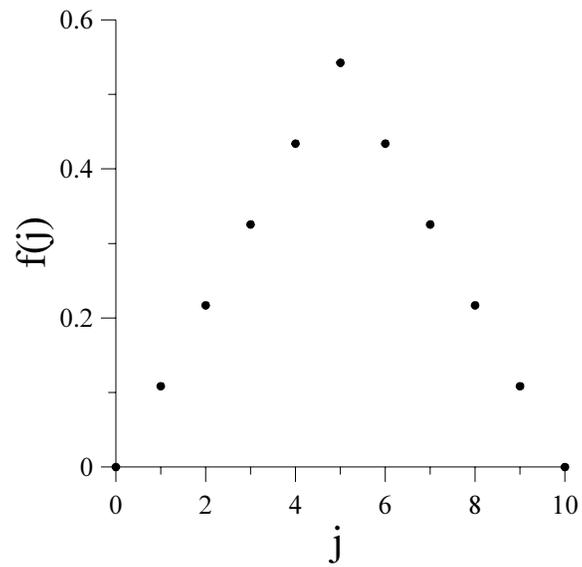

Fig. 3 Plot of the coefficients *f(j)* in the expansion of the entangled ancilla state for the simple linear case considered in the text, with $n=10$. The optimal choice for the *f(j)* coefficients gives 18% less error per quantum teleportation than is obtained from the linear case.